%% file: main.tex
\documentclass[%
twocolumn,
frontmatterverbose,
amsmath,amssymb,
aps,
superscriptaddress,
pra,
floatfix,
]{revtex4-2}

\usepackage[svgnames,psnames]{xcolor}
\usepackage{float}
\usepackage{url}
\usepackage{braket}
\usepackage{graphicx}% Include figure files
\graphicspath{{.}{figs/}{../figs/}}
\usepackage{afterpage}
\usepackage{dcolumn}% Align table columns on decimal point
\usepackage{bm}% bold math
\usepackage{tikz}
\usetikzlibrary{positioning,intersections,angles,quotes}
\usepackage[caption=false]{subfig}
\usepackage{comment}
\usepackage{physics}
\usepackage{mathtools}
\usepackage[colorlinks,citecolor=DarkGreen,linkcolor=FireBrick,linktocpage,unicode,hypertexnames=false]{hyperref}
\usepackage{xcolor}
\begin{document}

\title{Constructing Local Bases for a Deep Variational Quantum Eigensolver for
Molecular Systems}

\author{Luca Erhart}
\email{erhartl@hotmail.com}
\affiliation{%
  Graduate School of Engineering Science, Osaka University, 1-3 Machikaneyama, Toyonaka, Osaka 560-8531, Japan
}%

\author{Kosuke Mitarai}%
\email{mitarai@qc.ee.es.osaka-u.ac.jp}
\affiliation{%
  Graduate School of Engineering Science, Osaka University, 1-3 Machikaneyama, Toyonaka, Osaka 560-8531, Japan
}%
\affiliation{%
  Center for Quantum Information and Quantum Biology,
  Osaka University, 1-2 Machikaneyama, Toyonaka 560-8531, Japan
}%
\affiliation{%
  JST,
  PRESTO,
  4-1-8 Honcho, Kawaguchi, Saitama 332-0012, Japan
}%

\author{Wataru Mizukami}%
\email{wataru.mizukami.857@qiqb.osaka-u.ac.jp}
\affiliation{%
  Graduate School of Engineering Science, Osaka University, 1-3 Machikaneyama, Toyonaka, Osaka 560-8531, Japan
}%
\affiliation{%
  Center for Quantum Information and Quantum Biology,
  Osaka University, 1-2 Machikaneyama, Toyonaka 560-8531, Japan
}%
\affiliation{%
  JST,
  PRESTO,
  4-1-8 Honcho, Kawaguchi, Saitama 332-0012, Japan
}%

\author{Keisuke Fujii}%
\email{fujii@qc.ee.es.osaka-u.ac.jp}
\affiliation{%
  Graduate School of Engineering Science, Osaka University, 1-3 Machikaneyama, Toyonaka, Osaka 560-8531, Japan
}%
\affiliation{%
  Center for Quantum Information and Quantum Biology,
  Osaka University, 1-2 Machikaneyama, Toyonaka 560-8531, Japan
}%
\affiliation{%
  RIKEN Center for Quantum Computing (RQC),
  Hirosawa 2-1, Wako, Saitama 351-0198, Japan
}%

\date{\today}

\begin{abstract}
    Current quantum computers are limited in the number of qubits and coherence time, constraining the algorithms executable with sufficient fidelity. The Variational quantum eigensolver (VQE) is an algorithm to find an approximate ground state of a quantum system and expected to work on even such a device. The deep VQE [K. Fujii, \textit{et al.}, arXiv:2007.10917] is an extension of the original VQE algorithm, which takes a divide-and-conquer approach to relax the hardware requirement. While the deep VQE is successfully applied for spin models and periodic material, its validity on a molecule, where the Hamiltonian is highly non-local in the qubit basis, is still unexplored. Here, we discuss the performance of the deep VQE algorithm applied to quantum chemistry problems. Specifically, we examine different subspace forming methods and compare their accuracy and complexity on a ten H-atom treelike molecule as well as a 13 H-atom version. Additionally, we examine the performance on the natural occurring molecule retinal. This work also proposes multiple methods to lower the number of qubits required to calculate the ground state of a molecule.
   We find that the deep VQE can simulate the electron-correlation energy of the ground state to an error of below 1\%, thus helping us to reach chemical accuracy in some cases. The accuracy differences and qubits' reduction highlights the basis creation method's impact on the deep VQE. \\\\
   DOI: \href{https://dx.doi.org/10.1103/PhysRevApplied.18.064051}{10.1103/PhysRevApplied.18.064051}
\end{abstract}

\maketitle

\input{Sections/1_Introduction}
\input{Sections/2_Deep_VQE}

\input{Sections/3_Method}

\input{Sections/4_Numerical_simulation}

\input{Sections/5_Results}

\input{Sections/6_outlook}
\begin{acknowledgments}
L.E is supported by the Japanese Government (MEXT) scholarship for his PhD studies at Osaka University, Japan.
K.M. is supported by JST PRESTO Grant No. JPMJPR2019 and JSPS KAKENHI Grant No. 20K22330.
W.M. is supported by JST PRESTO Grant No. JPMJPR191A.
K.F. is supported by JST ERATO Grant No. JPMJER1601 and JST CREST Grant No. JPMJCR1673.
This work is supported by MEXT Quantum Leap Flagship Program (MEXTQLEAP) Grant No. JPMXS0118067394 and JPMXS0120319794.
We also acknowledge support from JST COI-NEXT program.
\end{acknowledgments}

\bibliography{main} % Include main.bib file

\appendix
\input{Sections/7_Appendix}
\end{document}

%% file: Sections/1_Introduction.tex
\section{Introduction}
Quantum algorithms have better asymptotic behavior than classical alternatives for some of today's most challenging and intriguing calculation problems. Outstanding examples are factorization of large numbers using the Shor algorithm \cite{Shor2006}, linear algebraic processes (matrix inversion)\cite{Harrow2009, Childs2017, Gilyen2019}, as well as promising results in quantum machine learning \cite{Liu2021}. Quantum chemistry will, however, arguably be the research field that gets impacted most by quantum computing \cite{Cao2019}. Simulating large molecules using a quantum computer will push our understanding of nature to other levels and significantly affect today's society. 
A specific intent in quantum chemistry is to calculate the electronic ground state of a molecule. Since the dimension of the Hilbert space grows exponentially to the system size, a molecular electronic structure can be exceedingly complex to solve on a classical computer.  
 
Despite recent developments \cite{Arute2019, Egan2020, Jurcevic2020}, we anticipate that quantum computers in the near future still have a limited number of qubits and only partial error resistance. Therefore the number of operations on a quantum device is limited, as the errors are building up to a point where we can not receive meaningful results. These restrictions confine us to the area of so-called noisy intermediate-scale quantum (NISQ) algorithms \cite{Preskill2018}. 

The variational quantum eigensolver (VQE) \cite{Peruzzo2014} algorithm is a promising approach for overcoming these obstacles. VQE is a method designed to find the ground state of a chemical system.
One uses a parameterized variational circuit, or ansatz, on a quantum device to prepare a trial state whose energy gets measured. A classical computer sets the parameters for the ansatz following some optimizer rules. 
By looping between the quantum device and a classical computer, one tries to minimize the energy. Using such an approach, it is possible to find a good approximation of the ground state.
The states created in such a way can be classically hard to represent. 
Depending on the ansatz, this only requires a circuit with a modest number of gates. There is, therefore, a possibility that VQE could be running on a NISQ device.

The system size solvable by the VQE is essentially limited by the number of qubits on the quantum device, making it challenging to apply it to large-scale systems. In classical quantum chemistry methods,  fragmentation techniques \cite{Gordon2011} proved to be highly successful in reducing the computational requirement to handle large-scale molecules. Recently divide-and-conquer techniques in quantum computing also got much attention to potentially solve the problem of the limited quantum resource \cite{Fujii2020, Mizuta2021, Zhang2021, Eddins2021}. They aim to decrease the number of qubits needed to solve complex problems. These techniques separate the original system into subsystems and solve them individually. The subsystem solutions set the starting point to formulate a meaningful result of the original problem in the next step. 

One such method is the deep VQE \cite{Fujii2020}.
It first separates the target quantum system into subsystems and obtains their approximate ground states by usual VQE.
The next step constructs a basis set for each subsystem by applying specific excitation operators to the subsystem ground states. The basis set is later used to form an effective Hamiltonian of the whole system.
Another VQE can then solve the effective Hamiltonian to obtain a ground state of the target system.
This process can be repeated multiple times to solve increasingly large systems.
The performance of the deep VQE has been analyzed for spin systems \cite{Fujii2020} and periodic materials \cite{Mizuta2021}, which have suggested that the deep VQE can produce accurate results while reducing the number of needed qubits simultaneously. However, only systems with minimal interactions between subsystems have been examined. Molecules are highly complex systems for which the Hamiltonian consists of many terms with various strengths. Such complex systems provide a unique set of challenges for the deep VQE. 
It is therefore vital to examine the performance of the deep VQE for such systems. 

In this work, we propose multiple methods to create the searchable subspace and compare their influence on the accuracy of the deep VQE using treelike molecules as a testbed. A 10-atom, and a 13-atom dendrimerlike molecule consisting only of hydrogen atoms, are chosen as the tree molecules in question. Additionally to these toy models we also apply the deep VQE to retinal to examine its performance on a natural molecule. Retinal is a natural occurring molecule of considerable size and is therefore a good indicator for the performance of deep VQE on a real complex molecule.
The strategy for creating a subsystem basis ultimately limits the deep VQE performance as it determines which subspace of the entire Hilbert space can be explored to express the ground state. To examine the impact of different bases on the deep VQE, we test three strategies to select excitation operators to form the subsystem basis. The first method followed the original paper and is based on the interaction operators. The second technique uses single-qubit Pauli operators to create a basis set, whereas the third obtains a basis using single-electron excitation and deexcitation operators.
Additionally to the different basis creation methods, we propose multiple techniques to control the needed qubits.
This work shows the remarkable accuracy and reduction of qubits the deep VQE offers for even complex quantum systems. 
We find that the deep VQE can simulate the electron-correlation energy of the ground state to an error of below 1\%, thus helping us reach chemical accuracy in some cases.
Our understanding of the various basis creation methods and the comparison of their performance provides an essential recipe for determining the bases of subsystems for the use of deep VQE in larger molecules of even more practical relevance. 

%% file: Sections/2_Deep_VQE.tex
\section{Theory}

\subsection{Deep VQE}
The deep VQE \cite{Fujii2020} is an algorithm of the divide and conquer family. Its purpose is to calculate the ground state of a quantum system. 
We review the algorithm of the original deep VQE, which can treat spin Hamiltonians consisting of 2-local interactions.
In deep VQE, we first divide the system into $M$ subsystems. The problem Hamiltonian with 2-local interactions can consequently be written in the form of
\begin{align}
    H = \sum_{i=1}^M H_i + \sum_{i,j=1}^M V_{ij},
\end{align}
where $H_i$ acts on subsystem $i$ and $V_{ij}$ on subsystems $i$ and $j$.
The interaction term $ V_{ij}$ decompose into operators $V_{\mu,i}$ acting only on single subsystems $i$:
\begin{equation}
	V_{ij} =\sum_\mu \lambda^\mu_{ij} V_{\mu,i} V_{\mu,j},
	\label{equ:interaction}
\end{equation}
where $\mu$ indicates the different interaction terms.
The deep VQE first finds ground states $\ket{G_i}$ of each subsystem Hamiltonian $H_i$ using a VQE algorithm.
Then we build a subsystem basis $\{\ket{b_{i,k}}\}_{k=1}^{K_i}$ for each subsystem $i$ with dimensions $K_i$ by acting with certain excitation operators $\mathcal{B}_i=\{B_{i,k}\}$ on $\ket{G_i}$.
The selection of the excitation operators $B_{i,k}$ is a crucial step which determines the performance of the deep VQE, and we will discuss it in detail in Sec. \ref{sec:method}.
To ensure the created basis set $\{\ket{b_{i,k}}\}_{k=1}^{K_i}$ is orthogonal a Gram-Schmidt orthogonalization is applied.

In the next step, we construct an effective Hamiltonian $H^{\mathrm{eff}}$ by projecting the original Hamiltonian $H$ to the subspace spanned by $\bigotimes_{i}\{\ket{b_{i,k}}\}_{k=1}^{K_i}$. 
This can be achieved by measuring all matrix elements $\bra{b_{i,k}}H_i\ket{b_{i,l}}$ and $\bra{b_{i,k}}\bra{b_{j,l}}V_{i,j}\ket{b_{i,p}}\ket{b_{j,q}} = 
\sum_{\mu} \lambda^\mu_{i,j} \bra{b_{i,k}} V_{\mu,i}\ket{b_{i,p}}\bra{b_{j,l}}V_{\mu,j}\ket{b_{j,q}}$.
Notice that calculating the expectation values only involves one subsystem, which can be calculated separately on a quantum computer. 
Now, we represent each subsystem by $N_i=\lceil\log_2 K_i\rceil$ qubits, and we again utilize the VQE using $N_{\mathrm{tot}} = \sum_i \lceil\log_2 K_i\rceil$ qubits to search for the ground state of $H^{\mathrm{eff}}$.
Even though $H^{\mathrm{eff}}$ whose dimension is $d=\prod_i K_i$ could, in principle, be mapped to  $\lceil\log_2 d \rceil \leqslant N_{\mathrm{tot}}$ qubits, it is beneficial to map each subsystem individually since it preserves the locality of the Hamiltonian.
The deep VQE algorithm can be repeated iteratively to treat larger and larger systems on a qubit-limited quantum computer.

\subsection{Deep VQE for fermionic system}
The central problem in quantum chemistry is finding the first-principles Hamiltonian's ground state, which describes interacting electrons. In second quantization, the Hamiltonian can be written in the form
\begin{equation}
\label{eq:fermionic_hamiltonian}
H = \sum_{o,p=1}^N h_{o,p} a_o^{\dagger} a_p +  \sum_{q,r,s,t=1}^N h_{q,r,s,t} a_q^{\dagger} a_r^{\dagger} a_s a_t.
\end{equation}
where $N$ is the number of orbitals of the molecule.
Here we consider how to apply the deep VQE to this Hamiltonian.

A frequent approach to treat the Hamiltonian of Eq. (\ref{eq:fermionic_hamiltonian}) on a quantum computer is to map the fermionic operators $a_o^\dagger$ and $a_o$ to qubit operators through, e.g., Jordan-Wigner transformation.
This leads to a Hamiltonian in the form of,
\begin{align}\label{eq:jw-transformed-hamiltonian}
    H = \sum_i h_i P_i,
\end{align}
where $P_i$ is a Pauli string, $P_i\in\{I,X,Y,Z\}^{\otimes N}$, and $h_i$ is a real coefficient.
The obstacle to apply the deep VQE is that $P_i$ is not 2-local but can be nonlocal when using JW transformation as we will see below.

After partitioning the system into $M$ subsystems and $M_{\mathrm{int}}$ interaction terms, we can write the Hamiltonian in Eq. (\ref{eq:jw-transformed-hamiltonian}) in the form of,
\begin{align}
    H = \sum_{i=1}^M H_i + \sum_{\mu=1}^{M_{\mathrm{int}}}\lambda^\mu V_{\mu, 1} \otimes V_{\mu,2} \otimes \cdots \otimes V_{\mu,M},
\end{align}
where $H_i$ and $V_{\mu,i}$ is an operator acting only on the $i$-th subsystem, and $\mu$ indexes different interaction terms.
To perform the deep VQE, we construct the effective Hamiltonian $H^{\mathrm{eff}}$ using the subsystem basis $\{\ket{b_{i,k}}\}_{k=1}^{K_i}$.
To do this, we measure the matrix elements $(H_i^{\mathrm{eff}})_{k,l} = \bra{b_{i,k}}H_i\ket{b_{i,l}}$ and $(V_{\mu, 1}^{\mathrm{eff}})_{k,l} = \bra{b_{i,k}}V_{\mu,i}\ket{b_{i,l}}$ for all combinations of $i$, $k$, $l$, and $\mu$.
Let $H_i^{\mathrm{eff}}$ and $V_{\mu,i}^{\mathrm{eff}}$ be $n_i=\lceil \log_2 K_i\rceil$-qubit operators with the evaluated matrix elements.
The total effective Hamiltonian can be written as,
\begin{align}\label{eq:effective_hamiltonian_global}
    H^{\mathrm{eff}} = \sum_{i=1}^M H_i^{\mathrm{eff}} + \sum_{\mu=1}^{M_{\mathrm{int}}} \lambda^\mu V_{\mu, 1}^{\mathrm{eff}} \otimes V_{\mu,2}^{\mathrm{eff}} \otimes \cdots \otimes V_{\mu,M}^{\mathrm{eff}}.
\end{align}
If we wish to perform the VQE of $H^{\mathrm{eff}}$, we must be able to efficiently measure the expectation value $\langle{H^{\mathrm{eff}}}\rangle = \bra{\psi} H^{\mathrm{eff}} \ket{\psi}$ for a given state $\ket{\psi}$ prepared on a quantum computer.
A usual approach to measure the expectation value of a Hamiltonian is to expand it into Pauli operators.
However, if we do so in Eq. (\ref{eq:effective_hamiltonian_global}), the number of Pauli operators grows exponentially to $M$ due to its nonlocality, which blocks efficient evaluation of $\langle{H^{\mathrm{eff}}}\rangle$.

We must avoid this exponential growth to apply the deep VQE to fermionic systems.
To this end, we first diagonalize $V_{\mu, i}^{\mathrm{eff}}$ classically and obtain unitary $U_{\mu,i}$ such that 
\begin{align}\label{eq:diagonalisation}
    V_{\mu, i}^{\mathrm{eff}} = U_{\mu,i}^\dagger\mathrm{diag}(\bm{\upsilon}_{\mu,i})U_{\mu,i},
\end{align}
where $\bm{\upsilon}_{\mu,i}$ are eigenvalues of $V_{\mu, i}^{\mathrm{eff}}$.
Note that this process can be performed in time $2^{O(n_i)}$ and that it is natural to assume $n_i$ is a small constant in the deep VQE. 
We can also construct a quantum circuit to realize a $2^{n_i}$-dimensional unitary $U_{\mu,i}$ in time $2^{O(n_i)}$.
Therefore, we can measure $\langle{V_{\mu, 1}^{\mathrm{eff}}\otimes\cdots\otimes V_{\mu, M}^{\mathrm{eff}}}\rangle$ for a given $\sum_{i=1}^{M}n_i$-qubit state $\ket{\psi}$ by first applying $U_{\mu, 1}\otimes\cdots\otimes U_{\mu, M}$ to $\ket{\psi}$ and then measuring it in the computational basis. Therefore, the required measurements to determine the expectation value of the interaction term scale linearly with the number of interactions in the system. 
Thereby, we can efficiently measure $\langle{V_{\mu, 1}^{\mathrm{eff}}\otimes\cdots\otimes V_{\mu, M}^{\mathrm{eff}}}\rangle$ and hence $\langle H^{\mathrm{eff}}\rangle$ as well.

Note that we prefer JW transformation for applying deep VQE to fermionic systems because it directly maps the $i$-th orbital to the $i$-th qubit; if the $i$-th qubit is $\ket{1}$, it indicates that an electron occupies the $i$-th orbital.
This property allows us to split systems naturally into subsystems using localized orbitals.
Although other techniques such as Bravyi-Kitaev (BK) transformation relaxes the locality of Pauli strings down to $O(\log N)$, it makes the correspondence between fermion occupation and qubit configurations not straightforward.
Note that the above technique may also be used for BK-transformation-based deep VQE to obtain the expectation values efficiently.

%% file: Sections/3_Method.tex
\section{Method} \label{sec:method}
\subsection{Basis creation strategies}
Choosing a basis set impacts the performance of the deep VQE. 
It ultimately determines the number of qubits needed to run the algorithm and the accuracy of the result. 
It is therefore vital to make an educated decision for the used method. 

Below we examine different procedures to generate the basis sets, which are summarized in Tab.~\ref{table_basis_set_rules} as well as their scaling in Table~\ref{basis_length_table}.
In the single Pauli method, we use the set of the Pauli operators $X, Y, Z$ acting on each qubit as $\mathcal{B}_i$.
In the interactions' approach, we apply all $V_{\mu,i}$ in Eq. (\ref{eq:effective_hamiltonian_global}) to the subsystem $i$ to create a basis, this leads to $\mathcal{B}_i=\{V_{\mu,i}\}_{\mu=1}^{M_{\mathrm{int}}}$.
For the particle conserving approach, we apply the excitation $a_s^\dagger $ or the de-excitation operators $a_s $ to every qubit $s$ of the $n_i$ qubits of the subsystem.
Additionally, to capture effects inside the subsystems, we applied swap gates between different qubits belonging to the same subsystem to create a basis. 
Therefore, in this approch, $\mathcal{B}_i = \{a_s\}_{s=1}^{n_i}\cup\{a_s^\dagger\}_{s=1}^{n_i}\cup\{\mathrm{SWAP}_{s,s'}\}_{s,s'=1}^{n_i}$.

We also use additional low-lying energy eigenstates $\ket{G^e_i}$ to generate the basis sets instead of using only $\ket{G_i}$ as has been done in the original paper \cite{Fujii2020}.
The superscript $e$ marks the $e$-th excited state and we define $\ket{G_i^0}:=\ket{G_i}$.
Note that if the subsystem Hamiltonian has degenerate ground states, $\ket{G^e_i}$ does not necessarily have different energy than $\ket{G_i^0}$. 
Such excited states can be constructed with an algorithm like the subspace-search VQE \cite{Nakanishi2018}.
After applying $\mathcal{B}_i$ to $\{\ket{G^e_i}\}_{e=1}^l$, we use the Gram-Schmidt algorithm to ensure that the created basis is orthogonal and minimal with $K_i$ elements.

\begin{table}
		\caption{Description of basis sets with starting vector and excitations. $n_i$ number of qubits in subsystem~$i$. $A_i$ number of qubits involved in strongest interaction. In Interactions fix qubits method $\varepsilon_{adapt}$ is selected such that the generated basis set is of fixed dimension. Otherwise a fixed $\varepsilon$ was chosen for the interaction-based methods.}
	\resizebox{0.5\textwidth}{!}{%
        \begin{tabular}{lll}
        \hline\hline
		\textbf{Method} & \textbf{Start vectors} & \textbf{Applied excitation} $B_{i,k}$ \\
		\hline
		\textbf{Interactions} & $ \ket{G^0_i} $& $\forall V_{\mu,i} \in V_{i,j,...} \quad \lambda^\mu > \varepsilon$\\
		\textbf{Interactions \& excited}  & $\ket{{G^e_i}}$ $\forall e < l $ & $\forall V_{\mu,i} \in V_{i,j,...} \quad \lambda^\mu > \varepsilon$ \\
		\textbf{Interactions fix qubits} & $ \ket{G^0_i} $& $\forall V_{\mu,i} \in V_{i,j,...}  \quad \lambda^\mu > \varepsilon_{adapt}$\\
		\textbf{Single Pauli }& $\ket{G^0_i}  $ & $P_s\quad s \in \{1,..., n_i\}$  \\
		\textbf{Single Pauli \& excited}& $  \ket{{G^e_i}}$ $ \forall e < l  $ & $P_s\quad s \in \{1,..., n_i\}$  \\
		\textbf{Single Pauli edge} & $\ket{G^0_i}  $ & $P_s\quad s \in A_i$\\
		\textbf{Particle conserving} & $\ket{G^0_i} $ &  SWAP\textsubscript{s,s'}, $a_s, a_s^\dagger \quad s,s'\in \{1,..., n_i\}$\\
		\textbf{Particle conserving edge} & $\ket{G^0_i} $ &  SWAP\textsubscript{s,s'}, $a_s, a_s^\dagger \quad s \in A_i$\\
		\hline
		\hline
	    \end{tabular}}
	\label{table_basis_set_rules}
\end{table}

The number of basis vectors determines how many qubits we need to run the deep VQE algorithm. 
We, therefore, also examine different approaches to reduce the dimensionality of the basis. 
For the interaction-based methods, we only consider interactions between subsystems with $\lambda^\mu$ exceeding a certain threshold $\varepsilon>0$. In the numerical experiments that follow, we examine the results for $\varepsilon = 10^{-2}$ if not stated otherwise. This selection was necessary since otherwise, the interaction-based methods would lead to basis sets that resemble no reduction from the entire Hilbert space. This was caused due to the numerous interaction terms between the subsystems for the molecular Hamiltonian.

The particle conserving and single Pauli methods as well can produce basis sets with a dimensionality that is challenging to represent on a quantum machine. We, therefore, also examine methods to truncate their basis sets. Our approach aimed to reduce the number of participating qubits in the subsystems for the basis-creating step leading to a subset of the entire basis. 
To do so, we select the interaction $\bigotimes_{i=1}^M \lambda^\mu V_{\mu,i}$ which has the largest absolute $\lambda^\mu$.
Let us denote the set of qubits involved in this interaction by $A_i$ for each subsystem $i$.
We then apply the corresponding excitation operators of the particle conserving or single Pauli methods to the qubits in $A_i$ to create the basis.
We call these methods particle conserving edges and single Pauli edges, respectively.
Mathematically, the former method uses $\mathcal{B}_i=\{a_s\}_{s\in A_i}\cup\{a_s^\dagger\}_{s\in A_i}\cup\{\mathrm{SWAP}_{s,s'}\}_{s,s'\in A_i}$ and the latter uses $\mathcal{B}_i = \{X_s,Y_s,Z_s\}_{s\in A_i}$.
The dimensionalities $K_i$ for the different strategies scale differently with increasing system sizes, which are summarized in Table~\ref{basis_length_table}.
These method will be compared numerically in Sec. \ref{sec:different_basis_numerics}.

The above strategies can reduce the number of qubits but in a hardware-agnostic way.
In practice, a quantum computer has a fixed number of qubits, and we wish to use as many qubits as possible within the hardware limitation.
With this in mind, we also propose an adaptive interaction-based strategy to create a basis set. 
We first order the interactions between the subsystems based on the interaction strength.
By gradually increasing the threshold $\varepsilon_{adapt}$, we find a basis set of the desired dimensionality.
This selection of thresholds can be performed for each subsystem independently.
We will present the numerical demonstration of this method in Sec. \ref{sec:fixed_qubit_numerics}.

\begin{table}[!h]
	\caption{Upper bound of the number of independent vectors $K_i$ in subsystem $i$. $n_i$ number of qubits in subsystem~$i$. $A_i$ number of qubits involved in strongest interaction.}
	{\setlength\tabcolsep{2pt}
		\begin{tabular}{lc}
			\hline\hline
			\textbf{Method} & \textbf{$K_i$} \\
			\hline
			\textbf{Interaction}     & $ \mathcal{O} (\text{\#Interactions}>\varepsilon)$ \\
			
			\textbf{Single Pauli}     &  $ \mathcal{O}(n_i)$ \\
			\textbf{Single Pauli edge}     &  $ \mathcal{O}(A_i)$ \\
			
			\textbf{Particle conserving} & $ \mathcal{O}(n_i^2) $   \\
			\textbf{Particle conserving edge} & $ \mathcal{O}(A_i^2) $   \\
			\hline\hline
	\end{tabular}}
	\label{basis_length_table}
\end{table}

\subsection{Remarks on degenerate eigenstates of subsystems} \label{sec:Degenerated states}
A subsystem Hamiltonian having degenerate ground states can lead to unexpected results.
A VQE then cannot find a unique ground state.
The found ground state would be semirandom and depends on multiple factors, such as the starting condition of the VQE and the noise in the quantum device.
In principle, all linear combinations of the degenerate ground states are possible solutions to the VQE without any treatment.
This results in an instability of the deep VQE since the random starting vectors, in general, lead to different basis sets and consequently to different $H^{\mathrm{eff}}$.

This dependency on the starting vector set is tricky as, without supplementary knowledge about the system, there is no justification for favoring one subvector set over the other without running it through the deep VQE first. 
Therefore, the deep VQE must consider all degenerate vectors as starting vectors to ensure that the result is independent of the randomness of the first VQE.
Alternatively, we should design the first VQE to return a unique solution for the reproducibility of the experiment by, for example, unfolding the degeneracy with constrained VQE \cite{Ryabinkin2018}.

%% file: Sections/4_Numerical_simulation.tex
\section{Numerical Simulation}

\subsection{Setup}
\subsubsection{Treelike molecules}
\begin{figure}%
    \centering
    \subfloat[\centering 10 Hydrogen atom treelike molecule ]{{\includegraphics[width=.2\textwidth]{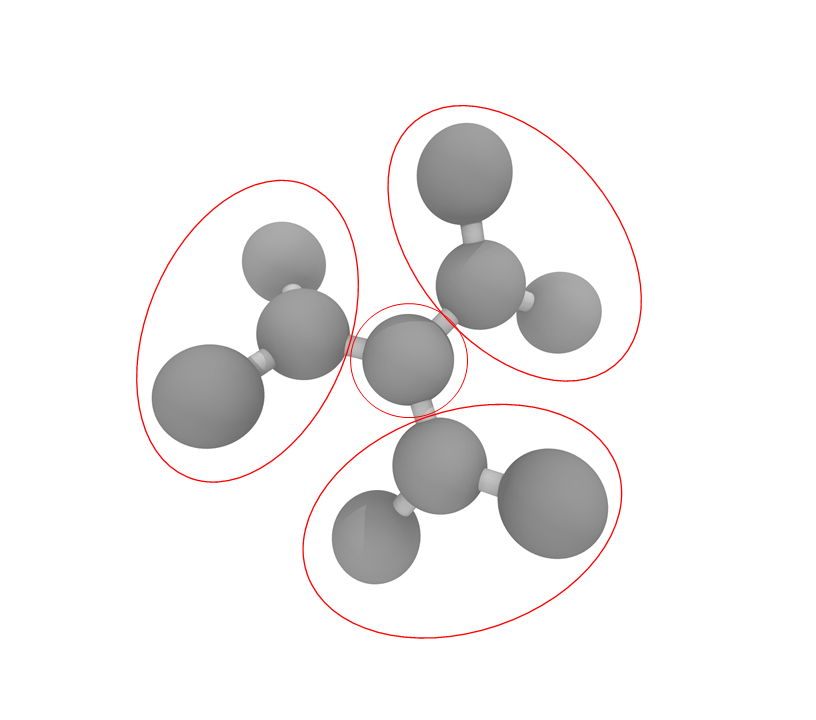}
    \label{fig:Tree_molecule_seperation_10}}}%
    \qquad
    \subfloat[\centering 13 Hydrogen atom treelike molecule ]{{\includegraphics[width=.2\textwidth]{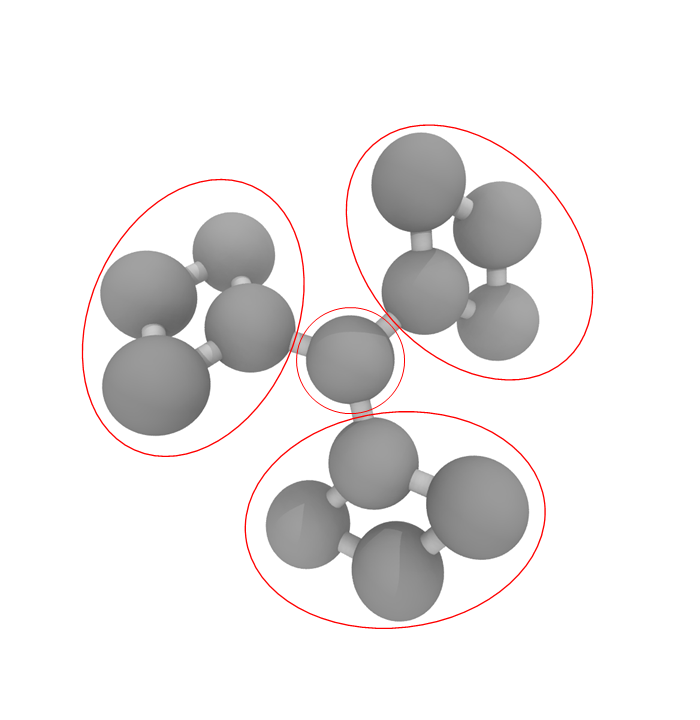}
    \label{fig:Tree_molecule_seperation_13}}}%
    \caption{Toy model molecules for deep VQE. Red circles indicate the partition of subsystems.\label{fig:Tree_molecule_seperation}}%
\end{figure}

We apply the deep VQE to two different toy models of treelike molecules in Fig. \ref{fig:Tree_molecule_seperation} a 10-atom and a 13-atom version. We describe the geometry of the treelike molecules in the Appendix \ref{sec:appendix}, notice that their geometry were not optimized rather it was chosen, to have an angle of 120 degrees and a torsions of 30 degrees.
Nakatani and Chan have used a similar molecule for benchmarking tree tensor network ~\cite{Nakatani2013}, however the distance between atoms were chosen different.
Since a treelike molecule naturally separates into different branchlike subsystems, it is also an ideal benchmark molecule for the deep VQE.
These molecules are the first step to calculating dendrimers or Cayley treelike molecules.
Today's industry frequently uses dendrimers, e.g., for pharmaceuticals \cite{Elham2021}.
Understanding dendrimers better would allow us to tailor them specifically to the applications.
We repeated the deep VQE simulation multiple times, enlarging the distance between atoms by applying a stretching factor ranging from 0.9 to 2.0 to all Cartesian coordinates, simulating the performance of deep VQE under the influence of different interaction strengths between subsystems.

Using STO-3G minimal basis set, the molecules in Figs. \ref{fig:Tree_molecule_seperation_10} and \ref{fig:Tree_molecule_seperation_13}, respectively, have 20 and 26 spin-orbitals.
The Hamiltonians of the molecules are obtained with PySCF \cite{Sun2018}.
We map it to a qubit system by JW transformation \cite{Jordan1993} implemented in OpenFermion \cite{McClean2020}, resulting in 20- and 26-qubit Hamiltonians.

For simulating the quantum states of these molecules on a classical machine, we use the python library Qulacs \cite{Suzuki2020}.

Figure \ref{fig:Tree_molecule_seperation} shows the separation of the treelike molecule into subsystems. Other selections of subsystems would be possible and an exciting research topic; however, we have to leave it to future research as it would exceed the scope of this work.
We use localized orbitals to define the subsystems based on the distance between their associated atoms.
The L\"{o}wdin orthogonalization method is employed to create well-localized orbitals. 

We use the full configuration interaction (FCI) to solve the molecules' ground state. The FCI solution is used as a reference for the performance of deep VQE. Additionally, we show the energy of the ``combined subsystem'' solution, which is the product state of all individual subsystem solutions, as well as the restricted Hartree-Fock solution. Hartree-Fock is a standard mean-field method that cannot account for the electron-correlation energy in the molecule.  

The performance of deep VQE depends partially on the subroutine VQE algorithm employed. However, this study aims to compare the different basis creation methods. Therefore we replace all VQE subroutines with direct diagonalizations of the matrix representation of the observable to find the ground state of the subsystem and the effective Hamiltonian. This replacement assures us that we find accurate eigenstates of the systems equivalent to having performed a FCI calculation. Additionally, the direct diagonalization replacement allows us to compare the different basis sets without the additional effects a realistic VQE algorithm would add, such as the noise of the quantum system. The ground state solution of the subsystems, as well as for the $H^{\mathrm{eff}}$ is determined using the exact diagonalization with SciPy \cite{Virtanen2020}. 

We find that the ground state of each subsystem of the 10-atom molecule [Fig. \ref{fig:Tree_molecule_seperation_10}] has a two-fold degeneracy for all stretching factors. We suspect that this degeneracy of the subsystems is due to a degeneracy of the spin eigenstates.
For the 13-atom tree molecule [Fig. \ref{fig:Tree_molecule_seperation_13}], each subsystem, except for the central hydrogen atom, has a unique ground state, but their first excited states are three-fold degenerate.
We also consider how this degeneracy affects the overall result in Sec. \ref{sec:degenerated-states-as-starting-vectors}.

The following subsections show the energy difference between the different basis sets and the FCI solution for the 10 and 13-atom tree molecules.  To better compare the different methods, we also show the weighted mean error of correlation energies over the different stretching factors and the required number of qubits to run the deep VQE on a quantum machine.
We define the weighted mean error of correlation energies as,
\begin{align}
     \parbox{9.5em}{Weighted~mean~error of~correlation~energies} = \frac{1}{|X|}  \sum_{x\in X}  \frac{E(x)-E_{\mathrm{FCI}}(x)}{E_{\mathrm{subsystems}}(x) -E_{\mathrm{FCI}}(x)},
\end{align}
where $E(x)$, $E_{\mathrm{FCI}}(x)$, and $E_{\mathrm{subsystems}}(x)$ are the energy obtained by the deep VQE using particular strategies, the FCI energy, and the energy of the "combined subsystem" solution at a stretching factor $x$. $X$ denotes the set of stretching factors for which we perform the calculations and $|X|$ indicates the of number of elements in $X$.
Here, $X$ is $\{0.9,1.0,1.1,1.2,1.3,1.4,2.0\}$.
We applied the weight $E_{\mathrm{subsystems}}(x) -E_{\mathrm{FCI}}(x)$ to the correlation energy error to take into account the difference in magnitude over the stretching factors. 

\subsubsection{Retinal}

\begin{figure}
    \includegraphics[width=0.4\textwidth]{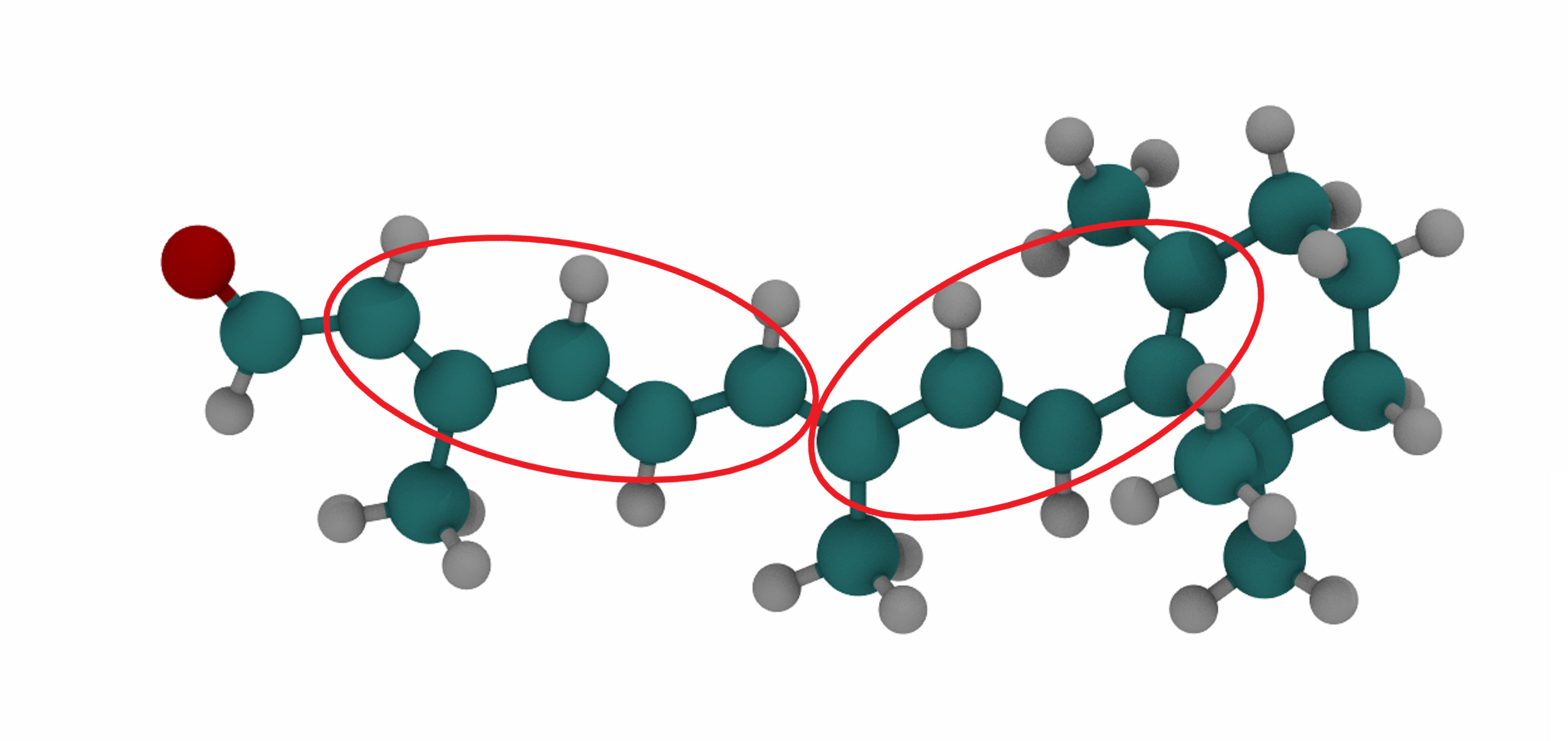}
    \caption{Retinal for deep VQE. Red circles indicate the partition of subsystems. Retinal was calculated using 20 $\pi$ orbitals splitting it equally into subsystems of ten orbitals. \label{fig:retinal2}}
\end{figure}

Additional to the toy treelike molecules, we also apply the deep VQE algorithm to a natural molecule, retinal. Retinal is essential in visual phototransduction, where visible light gets detected in our eyes.
We use library Gaussian 16 \cite{g16} at a B3LYP/6-31G** level of theory to find the optimized geometry of retinal. We show the geometry in the Appendix \ref{sec:appendix}. 
Retinal consists of 20 carbons, 28 Hydrogen, and one Oxygen. The number of orbitals in the STO-3G basis for the molecule forced us to calculate the ground state in an active space. We start by forming 20 $\pi$ orbitals using the PiOS \cite{Sayfutyarova2019} function of PySCF. Ten electrons are considered for this calculation. The $\pi$ orbitals take part in forming double bonds between the carbons in retinal. Then we localize the obtained $\pi$ orbitals using the Cholesky localisation method\cite{Aquilante2006}. 
The localised orbitals are used as the active space for molecule. As a reference for the deep VQE result we use a complete active space configuration interaction (CASCI) calculation. CASCI calculates the molecule's ground energy in the active space without optimizing the orbitals as CASSCF would do and is, therefore, less system-dependent. The Hamiltonian in this active space was mapped to a qubit system by a JW transformation using OpenFermion resulting in a 20 qubit Hamiltonian.
We split the molecule into 2 subsystems of each 10 spin orbitals as indicated in Fig.\ref{fig:retinal2}. Again this split is by no means unique and other separations could be examined. Both subsystems produced double degenerate eigenstates. This degeneracy most likely is due to a degeneracy of the spin eigenstates. In the case we consider only one starting vector we select the spin $\downarrow$ ground state as the stating vector. The results are shown in Sec. \ref{sec:retinal-result}. In the case of the edge methods, we used the bordering spin orbitals as the active qubits in the basis state finding method.

%% file: Sections/5_Results.tex
\subsection{Influence of starting vector for degenerate subsystems }\label{sec:degenerated-states-as-starting-vectors}

\begin{figure}
	\begin{center}
		\includegraphics[width=0.45\textwidth]{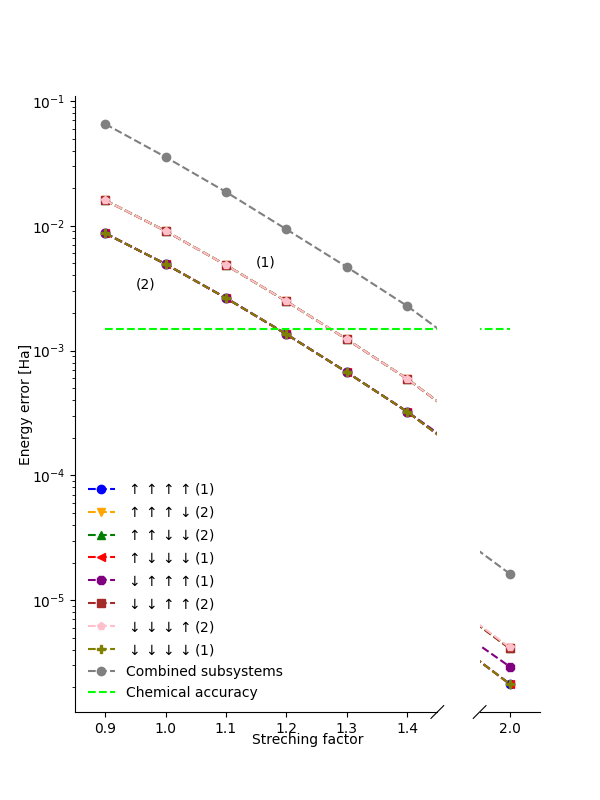}
		\caption{Performance of the particle conserving edges method with different incomplete starting vector compositions. The results are for the 10-atom tree molecule, which has a two degenerate ground state. The first marker describes the spin state of the central subsystem starting vector, whereas the three following markers indicate the spin state of the branchlike subsystems starting vectors. Since the results overlap in two lines, we introduce the labels (1) and (2) to indicate to which line the result belongs.}
		\label{fig:Results_flex_coef_10_atoms_spin_particle_conserving_edges}
		\includegraphics[width=0.45\textwidth]{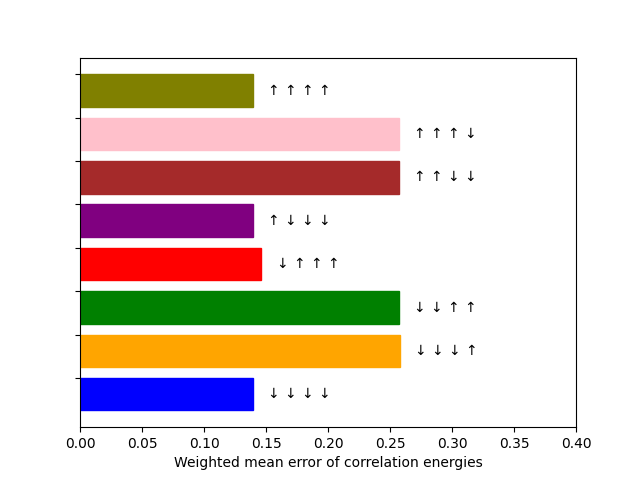}
		\caption{Weighted mean error of correlation energies for the particle conserving edges method with different incomplete starting vectors composition (1111).\label{fig:results__flex_coef_10_atoms_spin_particle_conserving_edges_mean_error}}
	\end{center}
\end{figure}

First, we discuss the dependency of the deep VQE on the starting vector. 
For the sake of readability, we only show the results using the particle conserving method to produce the basis in the deep VQE. 
We show the results for the 10-atom tree molecule in Fig. \ref{fig:Results_flex_coef_10_atoms_spin_particle_conserving_edges}. 
The ground states of all the subsystems of the 10-atom tree molecule are doubly degenerate.
The ground state found by the VQE or, in our case, the direct diagonalization is therefore not unique.
We use the spin state to distinguish the degenerate states.
We mark the spin-up ground state with $\uparrow$ and the spin-down ground state with $\downarrow$.
We only show a selection of all possible spin configurations due to redundancy resulting from the symmetry in the molecule.
We observe two distinct resulting energy levels, indicating that the stating vector $\ket{G_i}$ can considerably influence which Hilbert space can be searched for the overall ground state and, therefore, the performance of the deep VQE.

The choice of the starting vector is not trivial, and without further knowledge of the system, all possible starting vectors have to be considered. 
Additional starting vectors, however, come at the cost of requiring additional qubits to represent $H^{\mathrm{eff}}$.

\subsection{Different Basis methods}\label{sec:different_basis_numerics}
Next, we compare the performance of different basis methods with different starting vectors introduced in Sec.~\ref{sec:method} using the treelike molecules.
We compare two choices of starting vectors. The first choice uses single ground states of each subsystem, whereas the second uses additional eigenstates of each subsystem.
In the former case, we chose the spin configuration $\downarrow\downarrow\downarrow\downarrow$ for the starting vectors for 10-atom tree molecule since we had degenerate eigenstates. In the case of the 13-atom tree molecule, only the central subsystem had a twofold degeneracy, and we also chose the spin $\downarrow$ ground state as the starting vector.
For the interaction method, an $\epsilon=10^{-2}$ is chosen for the 10-atom tree molecule and an $\epsilon=10^{-3}$ for the 13-atom tree molecule. The number of considered starting vectors is indicated in the methods name in the bracket. The first digit indicates the number of starting vectors for the central subsystem, whereas the three following digits count the starting vectors for the branchlike subsystems.

Figures \ref{fig:all_methods_error_10_atom_distance} and \ref{fig:all_methods_mean_error_10_atom_distance} show the corresponding results for the 10-atom tree system.
We also show the number of qubits needed for each method in Table~\ref{tab:10_atom_molecule}.
We note that we are unable to perform the calculation for the particle conserving method with additional starting vectors due to its sizeable computational requirement.
The particle conserving methods with only one starting vector per subsystem performed exceptionally well in terms of accuracy.
For all tested basis creation methods, including an additional eigenstate significantly increased the accuracy of the deep VQE.
This is a consequence of the expanded Hilbert space that can be explored.

Unlike the other methods, we observe that the interaction methods with a fixed truncation have a decreasing accuracy with increasing stretching factors.
More concretely, the energy obtained by interaction methods jumps to the combined subsystem's energy at the stretching factor of 1.4.
The interaction methods also display different behavior regarding the saved qubits, showing a significant dependence on the stretching factor as shown in Table~\ref{tab:10_atom_molecule}.
This dependence is due to our selection process of the interactions to generate a basis set. 
We set a fixed cut-off of the interaction strength $\lambda^\mu$ for the interactions involved in the basis creation step. 
Consequently, as the interactions between subsystems become weaker by stretching the molecule, more interactions are disregarded.  
Lowering the dimensionality of the basis sets allows us to save more qubits but comes at the cost of lowering the method's accuracy.

\afterpage{%

\begin{figure}
\begin{center}
		\includegraphics[width=0.45\textwidth]{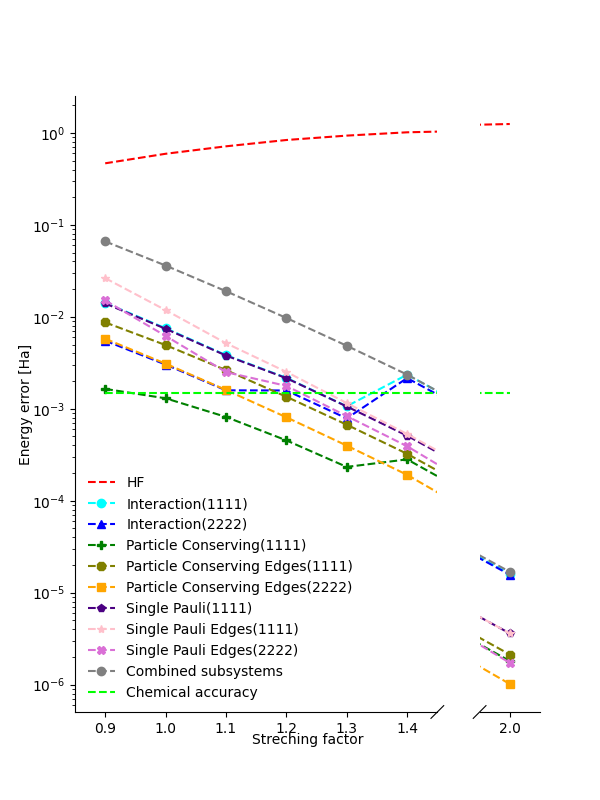}
        \caption{The energy difference between FCI and the deep VQE for the 10-atom tree molecule. We compare the different basis creation methods when using a single starting vector marked as (1111) and a complete degenerate basis set indicated by (2222). The first digit indicates the number of starting vectors for the central subsystem, whereas the three following digits count the starting vectors for the branchlike subsystems.
        \label{fig:all_methods_error_10_atom_distance}
	}
		\includegraphics[width=0.45\textwidth]{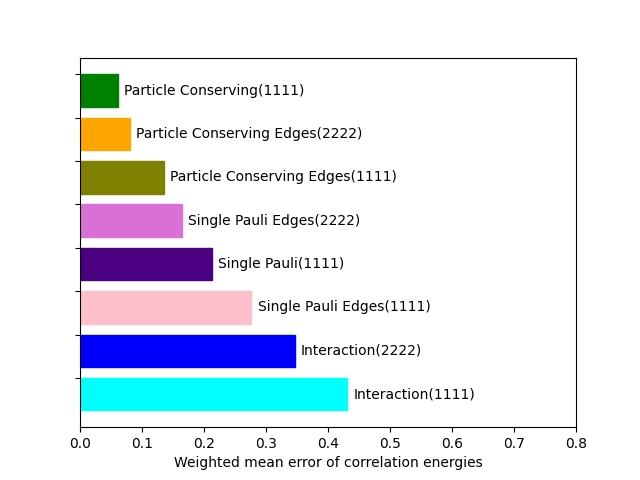}
		\caption{Weighted mean error of correlation energies for all methods for the 10-atom tree molecule\label{fig:all_methods_mean_error_10_atom_distance}.
	}
	\end{center}
\end{figure}
\begin{figure}
	\begin{center}
		\includegraphics[width=0.45\textwidth]{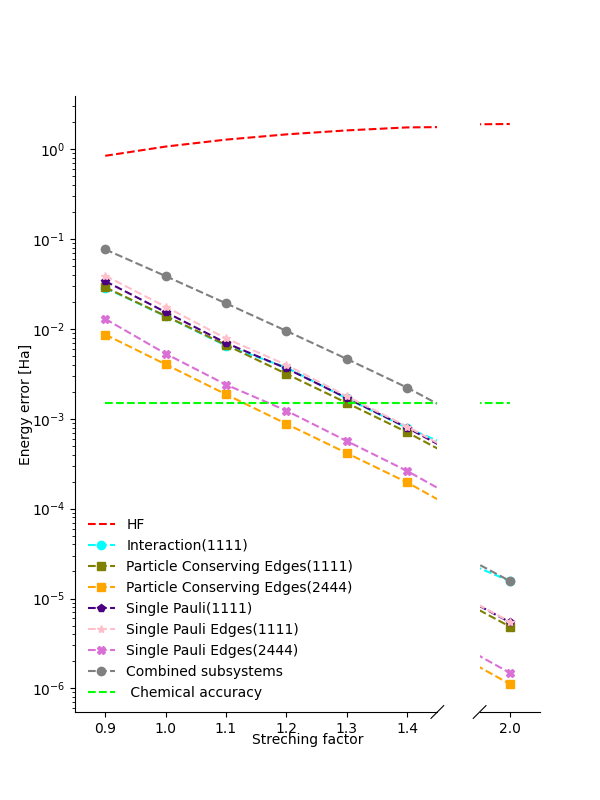}
        \caption{The energy difference between FCI and the deep VQE for the 13-atom tree molecule. We compare the different basis creation methods when using a single starting vector marked as (1111) and a full degenerate basis set indicated by (2444). The first digit indicates the number of starting vectors for the central subsystem, whereas the three following digits count the starting vectors for the branchlike subsystems.\label{fig:all_methods_error_13_atom_distance}}
		\includegraphics[width=0.45\textwidth]{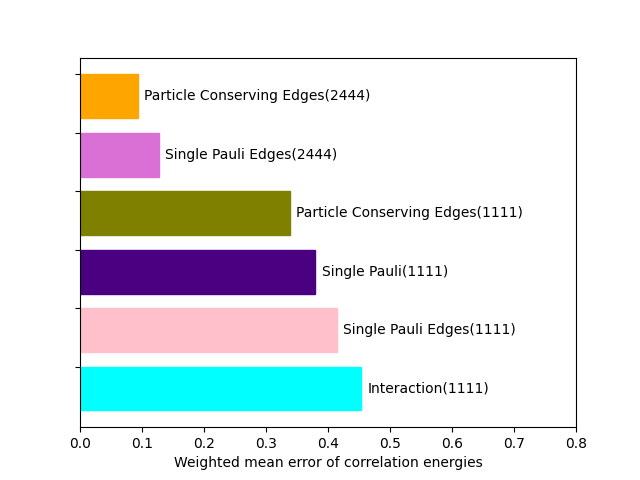}
        	\caption{Weighted mean error of correlation energies for all methods for the 13-atom tree molecule.}
        	\label{fig:all_methods_mean_error_13_atom_distance}
	\end{center}
\end{figure}

\begin{figure}
	\begin{center}
		\includegraphics[width=0.45\textwidth]{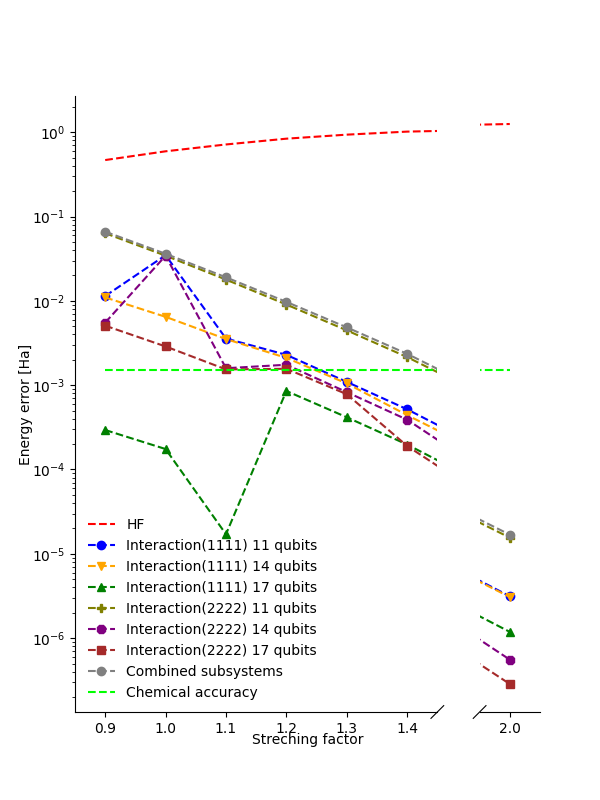}
        \caption{The energy difference between FCI and the deep VQE for the 10-atom tree molecule. We compare the interaction fixed methods with a different number of qubits. The fully degenerate basis set is indicated by (2222). The first digit indicates the number of starting vectors for the central subsystem, whereas the three following digits count the starting vectors for the branchlike subsystems.
        \label{fig:Results_flex_coef_10_atoms_compare_fix_qubits}}
		\includegraphics[width=0.45\textwidth]{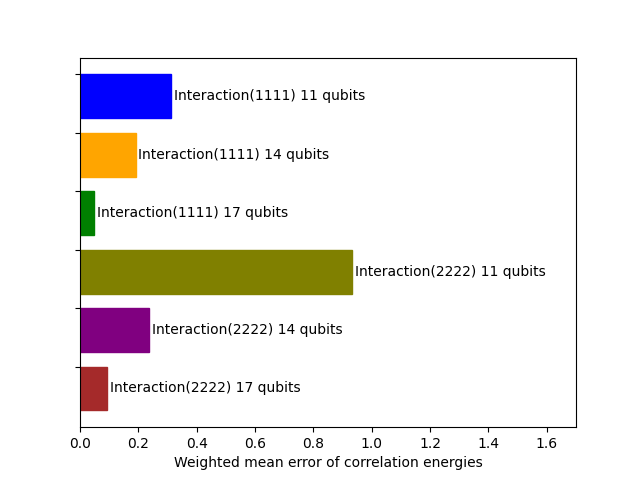}
			\caption{Weighted mean error of correlation energies for the interaction-based method with fixed number of qubits for the 10-atom tree molecule.\label{fig:Results_flex_coef_10_atoms_compare_fix_qubits_mean_error}}
	\end{center}
\end{figure}

\begin{figure}
	\begin{center}
		\includegraphics[width=0.45\textwidth]{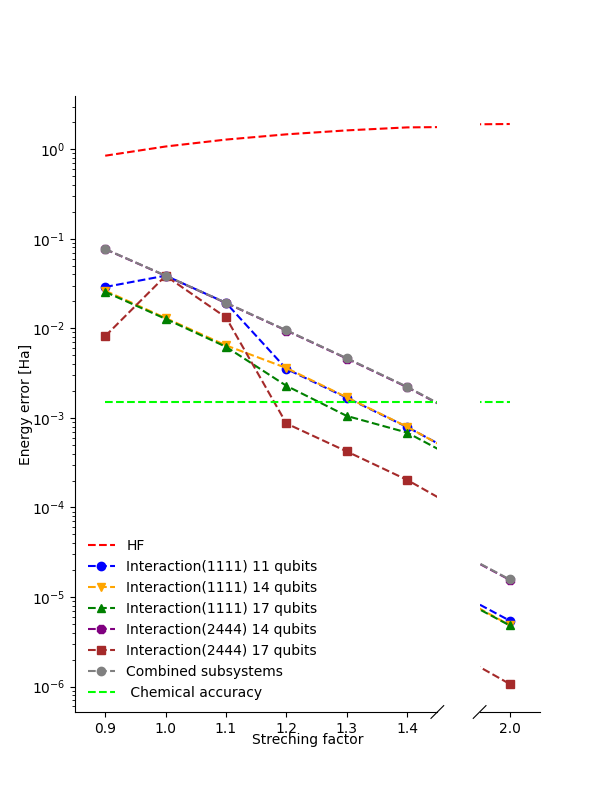}
		\caption{The energy difference between FCI and the deep VQE for the 13-atom tree molecule. We compare the interaction fixed methods with a different number of qubits. The fully degenerate basis set is indicated by (2444). The first digit indicates the number of starting vectors for the central subsystem, whereas the three following digits count the starting vectors for the branchlike subsystems.
		\label{fig:Results_flex_coef_13_atoms_compare_fix_qubits}}
		\includegraphics[width=0.45\textwidth]{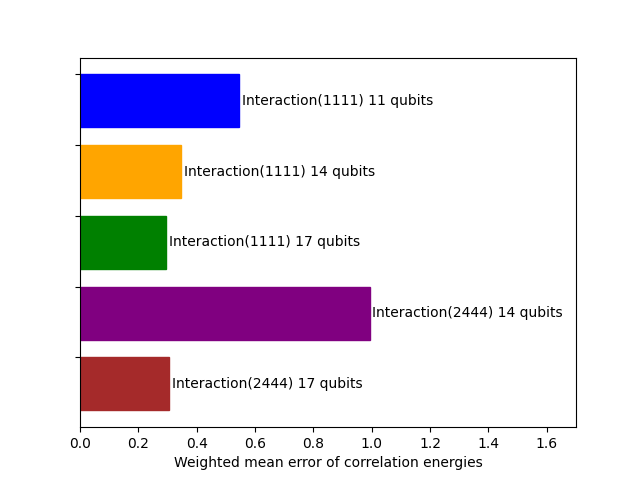}
				\caption{Weighted mean error of correlation energies for the interaction-based method with fixed number of qubits for the 13-atom tree molecule.\label{fig:Results_flex_coef_13_atoms_compare_fix_qubits_mean_error}}
	\end{center}
\end{figure}
\clearpage

}
Fig. \ref{fig:all_methods_error_13_atom_distance} and Fig. \ref{fig:all_methods_mean_error_13_atom_distance} show the results for the 13-atom tree molecule.
Here, only the central subsystem is doubly degenerate, and the first excited state of the branch subsystems for the 13-atom tree molecule is a triplet. 
To see the effect of adding additional excited states as stating vectors on the performance of the deep VQE, we consider all three states as additional starting vectors. 
Due to the high computational cost for the other methods, only the single Pauli edges and the particle conserving edges methods could be prepared with additional starting vectors. 
For both methods, the additional starting vectors are able to increase the accuracy of the deep VQE, but this comes at the expense of requiring extra qubits (see Table~\ref{tab:13_atom_molecule} for the exact number of qubits required for each method). 
The particle conserving edges methods performed slightly better than others.
This is consistent with the case of the 10-atom tree molecule, indicating an advantage of using the particle conserving approach in deep VQE for chemistry problems.

\begin{table*}
\caption{Qubits needed to represent the $H^{\mathrm{eff}}$ of the 10-atom tree molecule for the different basis creation methods. The number of considered starting vectors is indicated in the methods name in the bracket. The first digit indicates the number of starting vectors for the central subsystem, whereas the three following digits count the starting vectors for the branchlike subsystems. \label{tab:10_atom_molecule}}
\begin{tabular}{|l|c|c|c|c|c|c|c|}
\hline\hline
 \textbf{Stretching factor} &
  \textbf{0.9} &
  \textbf{1.0} &
  \textbf{1.1} &
  \textbf{1.2} &
  \textbf{1.3} &
  \textbf{1.4} &
  \textbf{2.0} \\
\hline
\textbf{Interaction(1111)($\lambda^\mu>10^2)$} &
\multicolumn{2}{|c|}{17} &
  14 &
  \multicolumn{3}{|c|}{11} &
  7 
  \\
  \hline
\textbf{Interaction(2222)($\lambda^\mu>10^2)$} &
  \multicolumn{2}{|c|}{17} &
  \multicolumn{3}{|c|}{14} &
  \multicolumn{2}{|c|}{10}\\
    \hline

\textbf{Particle conserving(1111)} &
  \multicolumn{7}{|c|}{17} 
 \\
   \hline
\textbf{Particle conserving edges(1111)} &
  \multicolumn{7}{|c|}{11} 
 \\
   \hline
\textbf{Particle conserving edges(2222)} &
  \multicolumn{7}{|c|}{14} 
 \\
   \hline
\textbf{Single Pauli(1111)} &
 \multicolumn{2}{|c|}{17} &
 \multicolumn{5}{|c|}{14}
 \\
    \hline
\textbf{Single Pauli edges(1111)} &
 \multicolumn{7}{|c|}{11}
\\
\hline
\textbf{Single Pauli edges(2222)} &
 \multicolumn{7}{|c|}{14}
\\
\hline\hline
\end{tabular}

\end{table*}

\begin{table*}
\caption{The numbers of qubits needed to represent the $H^{\mathrm{eff}}$ of the 13-atom tree molecule for the different basis creation methods. The number of considered starting vectors is indicated in the methods name in the bracket. The first digit indicates the number of starting vectors for the central subsystem, whereas the three following digits count the starting vectors for the branchlike subsystems. 
\label{tab:13_atom_molecule}}
\begin{tabular}{|l|c|c|c|c|c|c|c|}
\hline\hline
 \textbf{Stretching factor} &
  \textbf{0.9} &
  \textbf{1.0} &
  \textbf{1.1} &
  \textbf{1.2} &
  \textbf{1.3} &
  \textbf{1.4} &
  \textbf{2.0} \\
\hline
 \textbf{Interaction(1111)($\lambda^\mu>10^3)$}&
  \multicolumn{3}{|c|}{17} &
  \multicolumn{3}{|c|}{11} &
  7\\

\hline
 \textbf{Particle conserving edges(1111)}&
\multicolumn{7}{|c|}{11} 
\\
\hline
 \textbf{Particle conserving edges(2444)}&
\multicolumn{7}{|c|}{17} 
\\
\hline
\textbf{Single Pauli(1111)}&
\multicolumn{7}{|c|}{17} 
\\
\hline
 \textbf{Single Pauli edges(1111)}&
\multicolumn{7}{|c|}{11} 
\\
\hline
 \textbf{Single Pauli edges(2444)}&
\multicolumn{7}{|c|}{17} 
\\
\hline\hline
\end{tabular}
\end{table*}

\subsection{Interaction based deep VQE with fixed qubit numbers}\label{sec:fixed_qubit_numerics}
We show the results for the fixed qubit numbers approach in Figs. \ref{fig:Results_flex_coef_10_atoms_compare_fix_qubits} to \ref{fig:Results_flex_coef_13_atoms_compare_fix_qubits_mean_error}.
We fix the number of qubits to 11, 14, and 17.
For the 13-atom tree molecule, we could not produce an 11-qubit version as already including only the strongest interaction strength results in a basis set requiring more than 11 qubits.
The qubit number is increased by three per step to take into account that there are three equivalent branchlike subsystems of the molecules.

In Figs. \ref{fig:Results_flex_coef_10_atoms_compare_fix_qubits} and \ref{fig:Results_flex_coef_13_atoms_compare_fix_qubits}, we see that if we define a fixed number of qubits, we can avoid the decreasing accuracy for increasing stretching factors.
However, if the number of qubits is too restrictive, we cannot improve the result from the "combined subsystems". 
We see this behavior for the interaction(2222) method with 11 qubits for the 10-atom tree molecule in Fig. \ref{fig:Results_flex_coef_10_atoms_compare_fix_qubits}.

\subsection{Retinal}\label{sec:retinal-result}

Additional to the toy model of a 10 and 13-atom hydrogen trees, we also apply the deep VQE algorithm to retinal to examine the performance on a natural molecule. We apply different basis forming strategies and compared them based on their accuracy and the number of qubits in Tab. \ref{tab:retinal}. The calculation is performed with no stretching factors applied. The molecule got separated into two 10-qubit subsystems. With such a division, the eigenstates  of the subsystems are double degenerate. In the case when only one starting vector for the basis forming step was considered, we choose the spin-down starting vector. 
\begin{table}[h!]
\caption{The numbers of qubits needed to represent the $H^{\mathrm{eff}}$ of retinal for the different basis creation methods. The number of considered starting vectors is indicated in the methods name in the bracket. $\Delta$ E indicated the energy error of the method compared to the CASCI solution.\label{tab:retinal}}
\resizebox{0.5\textwidth}{!}{%
\begin{tabular}{|l|c|c|c|}
\hline\hline
 \textbf{Method} &
  \textbf{Qubits} &
  \textbf{Energy (H)} &
  \textbf{$\Delta$ E (mH)} 
 \\\hline
\textbf{Interaction(1,1) 12 qubits    }    & 12 & -838.2504 & 42.41 \\\hline
\textbf{Interaction(1,1) 14 qubits    }   & 14 & -838.2832 &9.578 \\\hline
\textbf{Interaction(2,2) 12 qubits     }   & 12 & -838.2911 & 1.759 \\\hline
\textbf{Interaction(2,2) 14 qubits   }      & 14 & -838.2924 & 0.362 \\\hline
\textbf{Particle Conserving (1,1) }          & 14 & -838.2821 & 10.75 \\\hline
\textbf{Particle Conserving Edges(10,10) }          & 12 & -838.2921 & 0.710 \\\hline
\textbf{Single Edges(8,8)    }                   & 12 & -838.2909 & 1.897 \\\hline\hline
\textbf{CASCI}                                                              &    & -838.2928 &             \\\hline
\textbf{HF}                                                                 &    & -838.1550 &           \\
\hline\hline
\end{tabular}}
\end{table}

Deep VQE proves to be effective in treating such a molecule as retinal. Especially the addition of additional states as starting vectors seems to be a valid strategy. For both the 12 and the 14 qubit interaction treatment, the addition of a second starting vector resulted in a significantly better ground state energy approximation. This can be a consequence of the doubly degenerate ground state of the individual subsystems. The particle conserving edges method with each 10 starting vectors performed again exceptionally well and is able to approximate the ground state energy in the STO-3G basis within chemical accuracy while saving 8 qubits. 

\subsection{Performance comparison}

Overall we achieve similar accuracy for the 10 and the 13-atom tree molecules (see Figs. \ref{fig:all_methods_error_10_atom_distance} and \ref{fig:all_methods_error_13_atom_distance}). We expect this from the minor influence the newly added outer atoms have on the overall ground state. This behavior would also explain the remarkable success the edge methods provide, saving up to 15 qubits with comparable accuracy to the other methods Tab.~\ref{tab:13_atom_molecule}. All methods are able to produce lower energies than the Hartree-Fock method. Multiple methods are able to approximate the ground state energies within an error of below 1\% of the electron correlation energy of the molecule. The electron correlation energy is equivalent to the error of the Hartree-Fock method shown in Figs. \ref{fig:all_methods_error_10_atom_distance} and \ref{fig:all_methods_error_13_atom_distance} and represents the error due to the mean-field approximation of Hartree-Fock. This improvement helped us reach chemical accuracy for some stretching factors.  The particle conserving edges method performed exceptionally well both for the treelike molecules as well for retinal. Especially the use of additional starting vectors proved to be beneficial for retinal. Using an edge method allows us to focus on changes affecting the orbitals involved in the most substantial interaction. We expect electrons occupying such orbitals to experience the most dramatic changes from the individual subsystem solutions when forming a bond with the other subsystems. To use the computational resources offered most effectively, it is advisable to use basis sets that focus on exploring the changes to the occupation of the most involved orbitals.

In contrast to the other methods, the interaction methods with a fixed truncation for the participating interactions significantly depended on the stretching factor. The number of qubits decreased with increasing distance. This reduction allows us to save more and more qubits but comes at the cost of decreasing accuracy. However, we consider this is not preferable if we have access to a quantum device with a fixed number of qubits. It seems unreasonable to not use all of them. The fixed-number qubit-interaction method we propose in this paper seems a more reasonable approach as it allows the use of all qubits. It also performs more stable for less interacting systems, not suffering from decreasing accuracy.

%% file: Sections/6_outlook.tex
\section{Conclusion and discussion}

The deep VQE approach successfully reduced the number of qubits to calculate a ground state of a complex chemical molecule. Notice that we do not exploit any symmetries in the molecules. Using such could be a further way to make deep VQE more efficient. All different basis creation methods we test could create a ground state energy within a few mHa of the FCI/CASCI solution. However, they show a significant difference in the accuracy and the number of qubits they could save. 
We discuss the challenges of degenerate subsystems and provid a solution in the form of additional stating vectors or marking the single starting vector by its spin state.
We also show that adding additional low-lying states as starting vectors for the basis creation method can improve the accuracy. However, the approximation of such states can be costly. 
Therefore, further research is needed to determine if this accuracy can be achieved using other basis creation methods. A possible approach would be to consider double Pauli excitation compared to the here used single Pauli method. Another exciting way to think about a new basis creation method is their similarity to VQE ansatzes. A new basis creation method could be created using a VQE ansatz with discrete fixed parameters.
We also propose methods to upper bound the number of qubits by using edge methods or selection of interaction in the interaction-based approach. These methods have proven effective strategies to reduce the number of needed qubits for the deep VQE. With these modifications to the deep VQE and a proper basis set, we believe that using deep VQE in a quantum chemistry setting can be highly beneficial.

%% file: Sections/7_Appendix.tex
\section{Geometry of molecule} \label{sec:appendix}
In Tabs. \ref{tab:geometry_10_atom}, \ref{tab:geometry_13_atom} and \ref{tab:geometry_retinal}, we show the geometry of the molecules used in this work. The distances are given in angstrom.

\begin{table}[!htb]
\caption{XYZ-coordinate of the 10 hydrogen molecule with stretching factor 1. \label{tab:geometry_10_atom}}
\begin{tabular}{llll}
& \textbf{X}& \textbf{Y}& \textbf{Z} \\
\textbf{H} & 0.000000  & 0.000000  & 0.000000\\
\textbf{H} & -1.732051 & -0.000000 & -1.000000                    \\
\textbf{H} & -1.848076 & -0.866025 & -2.799038                    \\
\textbf{H} & -3.348076 & 0.866025  & -0.200962                    \\
\textbf{H} & 0.000000  & 0.000000  & 2.000000                     \\
\textbf{H}  & -1.500000 & -0.866025 & 3.000000                     \\
\textbf{H}  & 1.500000  & 0.866025  & 3.000000                     \\
\textbf{H}  & 1.732051  & 0.000000  & -1.000000                    \\
\textbf{H}  & 3.348076  & -0.866025 & -0.200962                    \\
\textbf{H}  & 1.848076  & 0.866025  & -2.799038                   
\end{tabular}

\end{table}
\begin{table}[!htb]
\caption{XYZ-coordinate of the 13 hydrogen molecule with stretching factor 1.\label{tab:geometry_13_atom}}
\begin{tabular}{llll}
& \textbf{X}& \textbf{Y}& \textbf{Z} \\
\textbf{H} & 0.000000  & 0.000000  & 0.000000  \\
\textbf{H} & -1.732050 & -0.000000 & -1.000000                    \\
\textbf{H} & -1.848080 & -0.866030 & -2.799040                    \\
\textbf{H} & -3.348080 & 0.866030  & -0.200960                    \\
\textbf{H} & -3.464100 & -0.000000 & -2.000000                    \\
\textbf{H}& 0.000000  & 0.000000  & 2.000000                     \\
\textbf{H}& -1.500000 & -0.866030 & 3.000000                     \\
\textbf{H}& 1.500000  & 0.866030  & 3.000000                     \\
\textbf{H} & 0.000000  & 0.000000  & 4.000000                     \\
\textbf{H} & 1.732050  & 0.000000  & -1.000000                    \\
\textbf{H}  & 3.348080  & -0.866030 & -0.200960                    \\
\textbf{H} & 1.848080  & 0.866030  & -2.799040                    \\
\textbf{H}  & 3.464100  & 0.000000  & -2.000000                   
\end{tabular}
\end{table}

\begin{table}[!htb]
\caption{XYZ-coordinate of the retinal molecule.\label{tab:geometry_retinal}}
\begin{tabular}{llll}
& \textbf{X}& \textbf{Y}& \textbf{Z} \\
\textbf{O   }                & -8.60015                & 1.48468  & 0.81071  \\
\textbf{C    }               & 4.39241                 & -0.77397 & 0.51552  \\
\textbf{C}                   & 5.77904                 & -0.08948 & 0.48761  \\
\textbf{C }                  & 5.70665                 & 1.40739  & 0.77579  \\
\textbf{C  }                 & 3.33060                 & 0.08350  & -0.21654 \\
\textbf{C}                & 4.86219                 & 2.08607  & -0.30026 \\
\textbf{C}                   & 3.57180                 & 1.35840  & -0.61706 \\
\textbf{C}                   & 3.94256                 & -1.00034 & 1.97865  \\
\textbf{C}                   & 4.55288                 & -2.15570 & -0.16272 \\
\textbf{C}                   & 2.03748                 & -0.58705 & -0.44366 \\
\textbf{C}                   & 2.62992                 & 2.18732  & -1.45862 \\
\textbf{C}                   & 0.81416                 & -0.05758 & -0.20524 \\
\textbf{C}                   & -0.46332                & -0.72138 & -0.41518 \\
\textbf{C}                   & -0.46117                & -2.12317 & -0.97276 \\
\textbf{C}                   & -1.60447                & -0.03909 & -0.09459 \\
\textbf{C}                   & -2.96328                & -0.48737 & -0.21634 \\
\textbf{C}                   & -4.03235                & 0.27778  & 0.13513  \\
\textbf{C}                   & -5.42820                & -0.10540 & 0.04014  \\
\textbf{C}                   & -5.75579                & -1.47964 & -0.49730 \\
\textbf{C}                   & -6.37349                & 0.79473  & 0.44081  \\
\textbf{C}                   & -7.82162                & 0.62021  & 0.43151  \\
\textbf{H}                   & 6.22690                 & -0.23284 & -0.50558 \\
\textbf{H}                   & 6.43820                 & -0.59735 & 1.20232  \\
\textbf{H}                   & 6.71075                 & 1.84544  & 0.80511  \\
\textbf{H}                   & 5.26112                 & 1.58181  & 1.76272  \\
\textbf{H}                   & 5.44638                 & 2.18475  & -1.22944 \\
\textbf{H}                   & 4.61898                 & 3.11698  & -0.00678 \\
\textbf{H}                   & 4.64289                 & -1.66413 & 2.49891  \\
\textbf{H}                   & 3.88921                 & -0.05950 & 2.53474  \\
\textbf{H}                   & 2.94956                 & -1.45875 & 2.01686  \\
\textbf{H}                   & 5.40574                 & -2.68341 & 0.27827  \\
\textbf{H }                  & 3.67656                 & -2.79604 & -0.02741 \\
\textbf{H}                   & 4.74164                 & -2.05316 & -1.23676 \\
\textbf{H}                   & 2.08553                 & -1.61942 & -0.78134 \\
\textbf{H}                   & 2.14548                 & 2.97367  & -0.86405 \\
\textbf{H}                   & 1.84672                 & 1.59622  & -1.93374 \\
\textbf{H }                  & 3.19801                 & 2.70547  & -2.24209 \\
\textbf{H }                  & 0.75642                 & 0.95116  & 0.19869  \\
\textbf{H}                   & -1.46418                & -2.53053 & -1.09921 \\
\textbf{H}                   & 0.09232                 & -2.80140 & -0.31283 \\
\textbf{H}                   & 0.03720                 & -2.15163 & -1.94837 \\
\textbf{H}                   & -1.48044                & 0.96851  & 0.30163  \\
\textbf{H}                   & -3.13972                & -1.48470 & -0.60830 \\
\textbf{H}                   & -3.84165                & 1.27584  & 0.52734  \\
\textbf{H}                   & -5.29038                & -2.25362 & 0.12287  \\
\textbf{H}                   & -5.35887                & -1.59806 & -1.51154 \\
\textbf{H}                   & -6.82567                & -1.67873 & -0.53230 \\
\textbf{H}                   & -6.05111                & 1.76391  & 0.81603  \\
\textbf{H}                   & -8.21260                & -0.34930 & 0.05918 
\end{tabular}
\end{table}